\documentclass[12pt]{article} 
\pdfoutput=1 
\usepackage[utf8]{inputenc}
\usepackage{amsmath,amssymb,amsfonts} 
\usepackage[UKenglish]{babel} 
\usepackage[T1]{fontenc}
\usepackage{setspace}
\usepackage{physics}
\usepackage{hyperref}
\usepackage{xcolor}
\usepackage{graphicx}
\usepackage{lmodern}
\usepackage{booktabs}
\usepackage{ae,aecompl}
\usepackage{subfigure}
\usepackage[toc,page]{appendix}
\usepackage{epstopdf}
\usepackage{import} 

\usepackage{todonotes}
\usepackage{fullpage}
\allowdisplaybreaks 
\graphicspath{{figures/}}

\DeclareMathAlphabet{\mathpzc}{OT1}{pzc}{m}{it}

\def\be{\begin{equation}}
\def\ee{\end{equation}}

\title{Gravity and decoherence: \\the double slit experiment revisited} 
\author{Joseph Samuel$^{a}$\\ 
$^a$ Raman Research Institute, Bangalore 560 080}
\begin{document}

\maketitle

\begin{abstract}
The double slit experiment is iconic and widely used in classrooms to
demonstrate  the fundamental mystery of quantum physics. The puzzling
feature is that the probability of an electron arriving at the detector
when both slits are open is not the sum
of the probabilities when the
slits are open separately. The superposition principle of quantum mechanics
tells us to add amplitudes rather than probabilities and this
results in interference. This experiment defies
our classical intuition that the probabilities of exclusive events add.
In understanding the emergence of the classical world from the quantum
one, there have been suggestions by Feynman, Diosi and Penrose
that gravity is responsible
for suppressing interference.  This idea has been pursued in many different
forms ever since, predominantly within Newtonian approaches to gravity.
In this paper, we propose and theoretically analyse two `gedanken' or
thought experiments which lend strong support to the idea that gravity is
responsible for decoherence. The first makes the point that thermal
radiation can suppress interference. The second shows that in an accelerating
frame, Unruh radiation does the same. Invoking the Einstein
equivalence principle to relate acceleration to gravity, we
support the view that gravity is responsible for
decoherence.
\end{abstract} 

\section{ Gravity and Quantum Theory:}
The outstanding problem of theoretical physics today is the relation
between quantum theory and gravitation. Both these 
theories are experimentally very successful in their respective 
domains. Numerous experimental tests have vindicated Einstein's general
theory of relativity and the remarkable success of quantum physics in 
atomic, molecular, condensed matter physics and relativistic
quantum field theory needs no elaboration.  
The problem of merging these two successful theories into a coherent whole has
remained, despite much theoretical 
effort. 
A popular approach these days is to investigate new
theories which reduce, in the low energy limit,
to general relativity. Unfortunately, the energy scales of quantum
gravity are too high for us to get any experimental guidance in this venture.
The only guidance we have is from considerations of internal consistency
and aesthetics. Since aesthetic considerations are subjective, 
it is not entirely surprising that there is no 
consensus in the physics community today about the 
best approach to quantum gravity.

Faced with this situation, an alternative 
strategy is to understand the existing theories
better, by formulating {\it gedanken} or ``thought''
experiments in which both theories come into play. The ``thought'' experiments
do not actually have to be performed, though they must be performable 
in principle. 
As theorists, we can command imaginary resources beyond the reach of
current experiments, 
explore energy and length scales beyond the reach of technology 
and imagine idealised situations 
(like frictionless pulleys) which are not accessible to experimenters.
Gedanken experiments have been used in the past, most famously in
the Bohr-Einstein debates about the fundamentals of quantum mechanics.

In this paper, we propose two ``thought'' experiments, which are variations
of the double slit experiment, 
which Feynman \cite{feynman1965flp} described as ``{\it the} only mystery of
quantum mechanics''. While the electron 
double slit experiment has been performed in laboratories 
\cite{1976AmJPh..44..306M,doi:10.1119/1.16104,1367-2630-15-3-033018} 
over the world, 
the variations we propose here have not, to our knowledge, been  
discussed or analysed in any detail before. 

The idea that gravity decoheres the wave function has been championed by
Diosi \cite{diosi1987universal,diosi1989models} and Penrose\cite{penrose}. 
The line of thought can be traced
back even further to Feynman \cite{feynmanchapelhill,feynman2002feynman}.
One focuses on the large distance behaviour of quantum mechanics
rather than the short distance behaviour of gravity.
To quote Feynman \cite{feynmanchapelhill}, 
``I would
like to suggest that it is possible that quantum mechanics fails at large distances
and for large objects, it is not inconsistent with what we do know. If this failure
of quantum mechanics is connected with gravity, we might speculatively expect
this to happen for masses such that $GM^2 / c^2 = 1$, or $M$ near 
$10^{-5} {\rm grams}$.'' We will return to this quote
at the end of this paper.

The idea of gravity 
induced decoherence has been pursued in many forms.
See \cite{penrose2014,zych1,zych2,ADLER2016390,
diosi2015,minarPhysRevA.94.062111}
and references therein, for work on this topic. 
There are approaches which put in ``by hand'' a 
stochastic mechanism that effects non Unitary evolution, thereby altering the quantum theory
to include a description of the ``collapse of the wave function''.
This is not the approach we take here. We wish to keep both GR and Quantum theory intact and
look for decohering effects that destroy superpositions on larger scales.

Our objective here is to 
propose two gedanken experiments E1 and E2 and analyse them mathematically
to work out the expected outcome, 
{\it using only known physics}. 
The two experiments are very similar
in that they are both double slit experiments. Both experiments are performed
under stationary conditions, with a monoenergetic electron beam tuned in intensity 
so that there is just about one electron at any time in the apparatus. 
E1 considers the electron double slit experiment in a thermal photon bath: 
we find that thermal fluctuations of the electromagnetic field destroy
coherence of the electron beams. 
E2 considers the double slit experiment in a uniformly accelerated frame. 
We find that here too coherence
is destroyed by fluctuations, though now they are quantum fluctuations of the Minkowski vacuum, seen by the accelerated Rindler observer 
as thermal. Our objective in linking these two experiments is that the first (E1) is based on very familiar laboratory physics,
which will be readily accepted by the reader.  The second (E2) is far removed from everyday experience. Yet, the mathematical
analysis we present for E1 and E2 is  
virtually identical and serves as a bridge 
connecting everday physics to exotic physics.

\section{E1:Double Slit experiment in a thermal environment}
\begin{figure}[h!]
\centering
\includegraphics[scale=0.6]{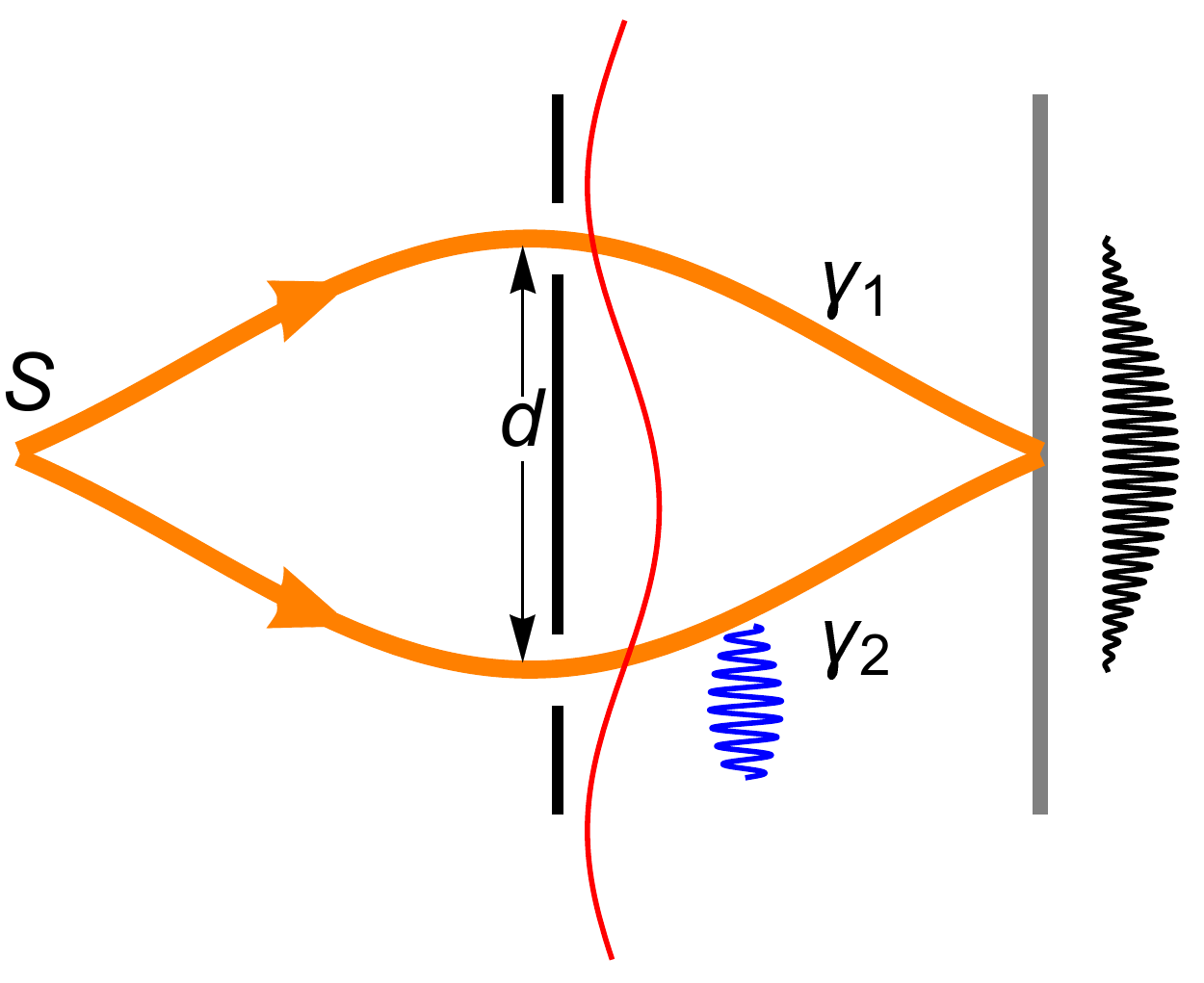}
\caption{(Color online) The Double Slit Experiment: Figure shows a schematic
diagram of the double slit experiment. Monoenergetic 
electrons emerge from a source at $S$, 
pass through two slits, separated by a distance $d$ and
fall onto a detector, where an interference pattern 
(black squiggly line on extreme right)
can be observed.
In the presence of thermal radiation, the electrons 
can scatter off the ambient photons 
(shown as a red wavy line and a blue squiggly line) and 
the interference pattern is destroyed.
}
\end{figure}
Figure 1 shows the setup for the double slit experiment. The source S 
emits  a beam of monoenergetic, single electrons which are allowed
to pass through two slits (separated by a distance $d$) in a screen, (shown in black) and fall on a detector (shown in grey). The interference
pattern expected is drawn just behind the detector. Shown in the figure (Fig.1) are two paths $\gamma_1$ and $\gamma_2$
which the electron could have taken to get from the source to the detector. The probability for arriving at the detector
with both slits open is $P_{12}$, which is not the same as $P_1+P_2$, the sum of the probabilites $P_1$ and $P_2$ of arrival with the slits open
one at a time. The difference $I=P_{12}-P_1-P_2$ is the quantum interference term. 

Let us clarify here that the two paths  $\gamma_1$ and $\gamma_2$ can be arbitrary 
curves along which the electrons are guided by external potentials. We must specify that the kinetic energy 
of the electrons
is much smaller than their rest energy $m_e c^2$ {\it. i.e},
the electrons are moving at non-relativistic speeds.  
Else, there is a possibility of pair creation under 
the influence of the external potentials and slits, 
which could confuse the experiment.

Let us now look at the effect of a thermal environment on this experiment. 
We make the idealised assumption that the entire apparatus is 
transparent to photons. 
If the experiment is done at a finite temperature $T$ (much less than $m_ec^2/k_B$), 
there will be
ambient black body photons present. These photons could scatter off the electrons and in doing so, impart some momentum to them. The ambient
photons are shown as wavy lines (Fig.1) in blue and red. The red photons have long wavelengths (long compared to $d$, the slit separation) and these do not carry
much momentum. The blue photons have shorter wavelengths than $d$ and 
so have enough momentum to deflect an electron from a bright fringe into a dark one.
At a temperature $T$, there is an abundance
of thermal photons at a frequency $\nu=\frac{k_B T}{2\pi \hbar}$, but higher frequency photons are exponentially scarce.
We would expect then, that the interference pattern is progressively washed out as the temperature is raised. 
This physical argument shows that
as $T$ increases beyond $\hbar c/(k_Bd)$ the electron interference pattern 
disappears and we recover the classical probability rule. Put differently, the 
thermal electromagnetic field has spatial correlations that die out with distance as $\exp{[-(x-x')/\lambda_w]}$ where $\lambda_w=\frac{\hbar c}{k_BT}$ is the Wien
wavelength. At high temperatures, the electromagnetic field fluctuations over the two slits are independent and the interference pattern is destroyed.
(By high temperatures, we mean $\hbar c/d<k_B T<<m_e c^2$, where $m_e$ is the electron mass, else thermal production of electron positron pairs would 
confuse the experiment.)

This physical argument can be made mathematically precise. In the absence of the electromagnetic field, let the amplitude 
for arriving at the detector via path $\gamma_1$ be $\Psi_1$ and similarly $\Psi_2$ 
the amplitude for arrive via path $\gamma_2$. For simplicity, we will assume that $|\Psi_1|=|\Psi_2|$.
Then $P_1=|\Psi_1|^2$, $P_2=|\Psi_2|^2$ and $P_{12}=|\Psi_1+\Psi_2|^2$ . The interference
term is
\begin{equation}
I=\Psi_2^* \,\,\Psi_1 +\Psi_1^* \,\,\Psi_2 
\label{interference0}
\end{equation}
and the fringe visibility is unity.
In the presence of the electromagnetic field, these amplitudes are modified
to $[\exp{ie/(\hbar c) \int_{\gamma_1} \bf{A}.d\bf{x}}]\Psi_1$ and $[\exp{ie/(\hbar c) \int_{\gamma_2} \bf{A}.d\bf{x}}]\Psi_2$,
where $\bf{A}$ is the vector potential of the electromagnetic field. 
Eq. (\ref{interference0}) naturally brings in the closed
Wilson loop $ {\cal W}(\gamma)=[\exp{ie/(\hbar c) \int_{\gamma} \bf{A}.d\bf{x}}]$ where 
the loop $\gamma=\gamma_1+\overline\gamma_2$, 
goes from source to detector via $\gamma_1$ following the arrow (Fig.1)
and returns via $\gamma_2$ {\it against} the arrow.
The Wilson loop measures the total magnetic flux passing through the loop $\gamma$ and
puts an additional random phase into the interference term. 
The interference term is now given by
\begin{equation}
I=<{\cal W}>\Psi_2^* \,\,\Psi_1 \,\,+<{\cal W}^\dagger>\Psi_1^* \,\,\Psi_2 
\label{interference}
\end{equation}
In order to calculate this quantity, we decompose the electromagnetic field $\bf{A}(\bf{x})$ into modes ${\bf u}_l(\bf{x})$, 
\begin{equation}
{\bf A}({\bf x})= \sum_l [{\bf u}_l({\bf x})\, a_l+{\bf u}_l^*({\bf x})\,a_l^\dagger]
\label{modexpansion}
\end{equation}
where $l$ is a label for the modes.
In E1, the modes are labelled by the momentum and polarisation, $l=\{\bf{k},\lambda\}$ and ${\bf u}_{\bf{k} \lambda}=
\frac{{\bf \epsilon}_{{\bf k},\lambda}}{\sqrt{2 V \omega_{\bf{k}}}}\exp{i\bf{k}.\bf{x}}$,
where $\omega_{\bf k}=|\bf{k}|$ is the frequency and $V$ the volume of the box. (We use periodic boundary conditions, so space is a torus).
Computing the exponent of the Wilson loop, we find
$\int_\gamma {\bf A}({\bf x}).d{\bf x})=\sum_l (a_l \alpha_l+ a_l^\dagger
\alpha^*_l)$.
Here $\alpha_l$, the ``form factor'' of the loop $\gamma$ is given by
\begin{equation}
\alpha_l=\int_{\gamma} {\bf u}_l({\bf x}).d{\bf x}.
\label{alpha}
\end{equation}

We then find that the Wilson loop expectation value can be
written as a product of independent contributions from the individual modes: 
\begin{equation}
<{\cal W}>=\prod_{l} <{\cal W}_l>
\label{wilsonproduct}
\end{equation}
Since each mode is an oscillator, the contribution from each mode
can be worked out (see the appendix for mathematical  details). 
${\cal W}_l$ is a Unitary operator 
$\exp{i [a_l \alpha_l+a_l^\dagger \alpha^*_l}]$, 
where $a_l$ destroys and $a_l^\dagger$ creates a photon in 
the $l {\rm th}$ mode. $\alpha_l$ is the ``form factor'' 
of the loop $\gamma$, essentially, the Fourier transform of the loop. 
To find the expectation value of ${\cal W}_l$,
we use the thermal average
 \begin{equation}
<{\cal W}_l>=\frac{{\rm{Tr}}[{\cal W}_l \exp{- H_l/(k_B T)}]}{{\rm{Tr}} [\exp{-H_l/(k_B T)}]} 
\label{boltzmann}
\end{equation} 
where $H_l$ is the oscillator Hamiltonian for the $l{\rm th}$ mode. 
Each $<{\cal W}_l>$ is real ($<{\cal W}_l>=<{\cal W}_l>^*=<{\cal W}^\dagger_l>$ and lies between $0$ and $1$. 
$<{\cal W}_l>$ quantifies the decohering
effect of the single mode $l$.

The product (\ref{wilsonproduct})
of the $<{\cal W}_l>$, which measures the total decohering effect of all modes is also real and lies between 
$0$ and $1$. 
Since $<{\cal W}>$ multiplies the interference term,
the fringe visibility is the thermal average  $<{\cal W}>$ of the Wilson loop
${\cal W}$ (we drop the label $\gamma$ since the loop is held fixed).
Our analysis (described in detail in the appendix) 
yields the closed analytic form 
\begin{equation}
<{\cal W}>=\exp{[-\frac{e^2}{2\hbar c} \sum_{l}  (|\alpha_l|^2 \coth{\frac{\hbar \omega_l}{2 k_B T})}]}
\label{finalresult}
\end{equation} 
This result is valid for arbitrary closed loops $\gamma$, where $\alpha_l$ is the Fourier transform of the loop. 
For ease of calculation, we choose $\gamma$ to be a square of side $d$. (Such a loop could be realised
in a an interferometer.)
The form of $<{\cal W}>$ for this specific choice of $\gamma$ 
is plotted in Fig.2 as a function of $\frac{dk_BT}{\hbar c}$.
\begin{figure}
\includegraphics[height=8.0cm,width=10.5cm,scale=5]{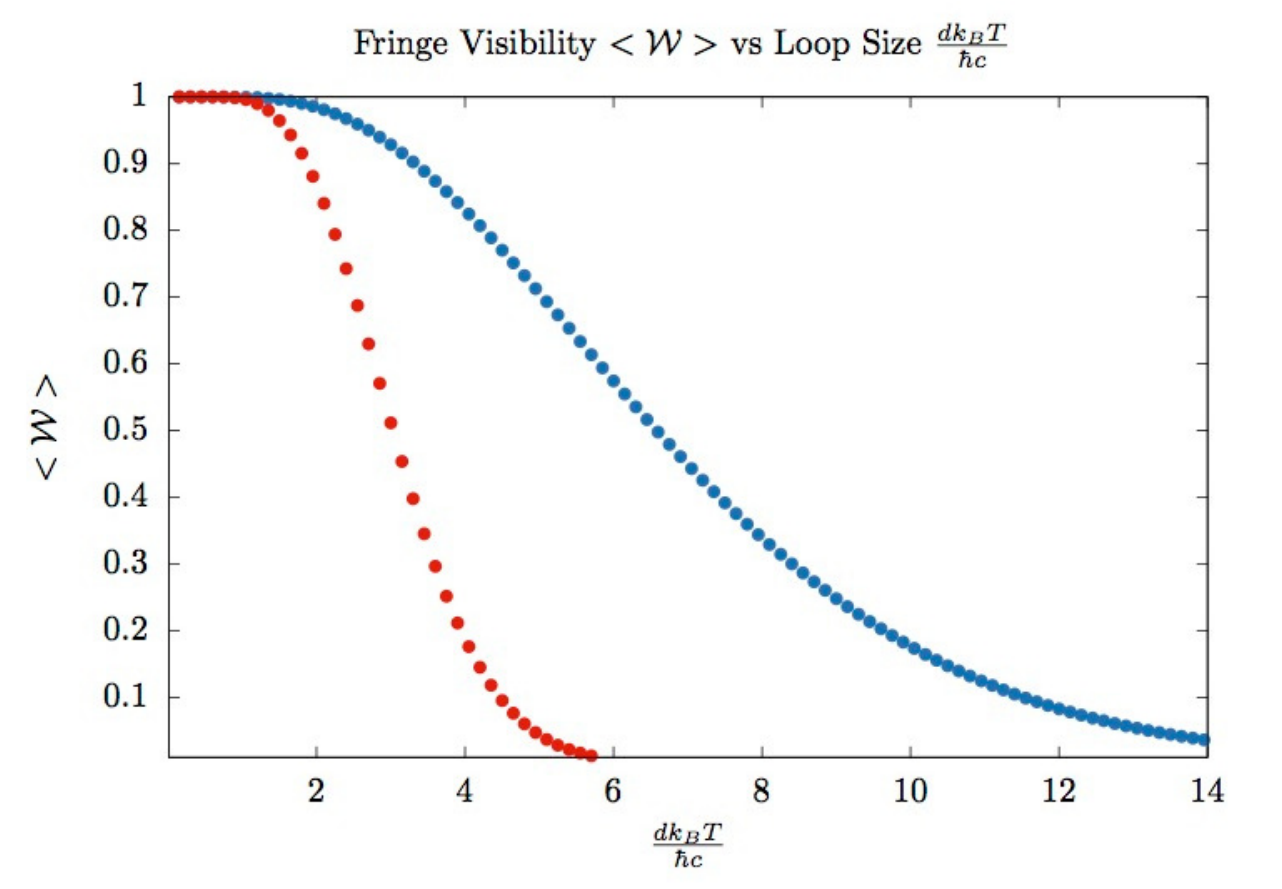}
\caption{(Color online) The loss of coherence with size and temperature: 
Figure shows a plot of $<{\cal W}>$ versus $\frac{d k_B T}{\hbar c}$
for two values of the coupling constant: $\frac{e^2}{\hbar c}=1/137$ (upper curve, in blue)
and $9/137$ (lower curve, in red). The first number is relevant to electrons and the second
to triply charged ions. Note that the coherence decreases from unity  
at zero temperature and size to zero at large temperatures and sizes. 
}
\end{figure}
As expected, $<{\cal W}>$ is unity at low temperatures (and small loops) and decreases to zero at higher temperatures (and larger loops). 
Thus, the interference pattern is washed out by thermal effects.  This calculation confirms the physical picture given
earlier in terms of photons.

\section{ E2 :Double Slit Experiment in an Accelerated Frame:}
Let us now consider our second thought experiment E2, performing the double slit experiment in a Rindler frame, 
which is a uniformly accelerated frame. 
We suppose that the apparatus, at rest in the accelerating frame, is transparent to photons
and that electron-positron pair creation effects can be neglected. 

\begin{figure}
\includegraphics[height=7.5cm,width=8.5cm,scale=0.6]{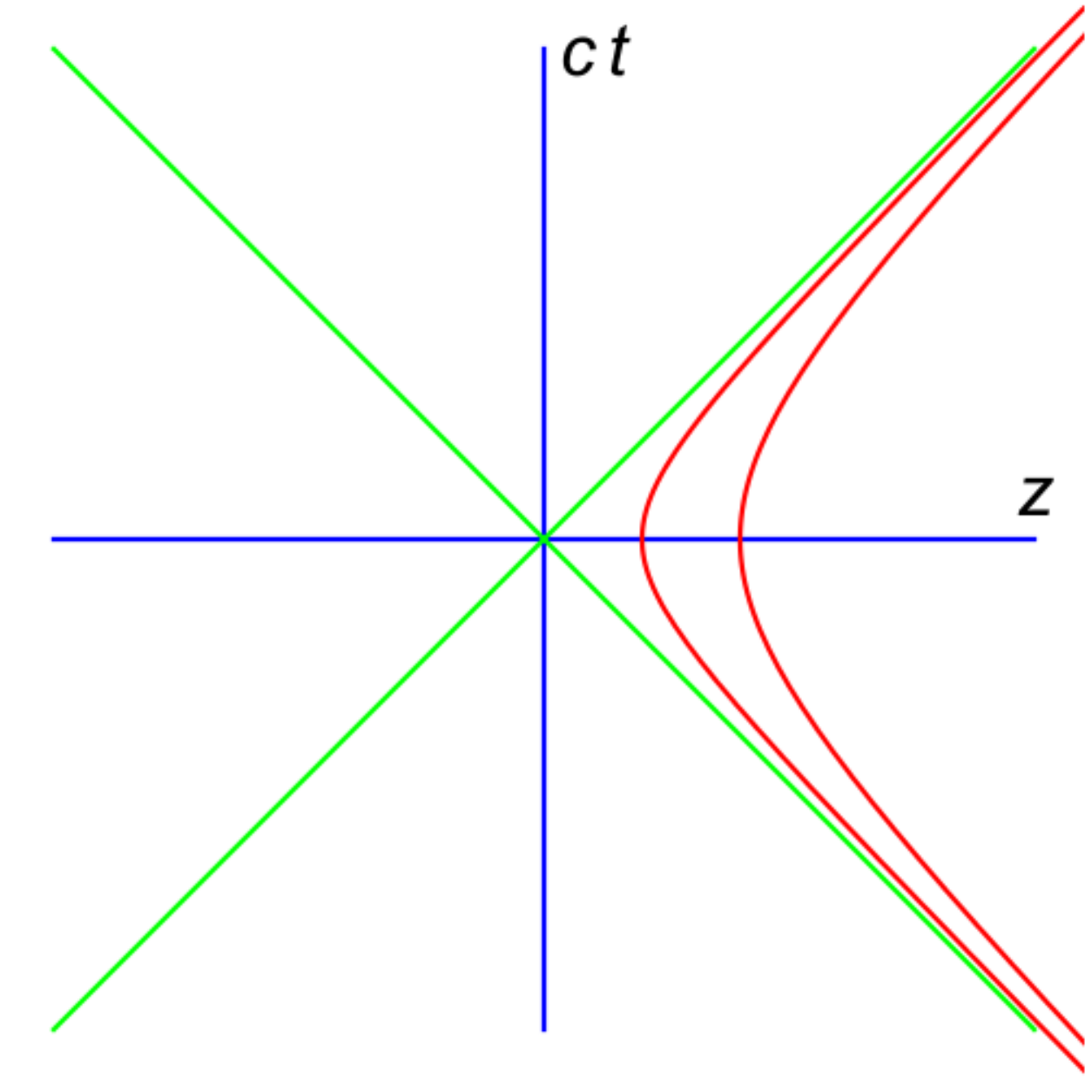}
\caption{(Color online) Uniformly accelerated observers: 
Figure shows Minkowski space and the world lines of uniformly
accelerated observers (curved lines in red). 
These observers are confined to the right Rindler wedge and
see a thermal background of radiation. Also shown 
(green diagonal straight lines ) are the 
light cones and (straight lines in blue) 
the coordinate axes of the inertial observer. 
}
\end{figure}
Fig. 3 shows empty Minkowski spacetime and the world lines of uniformly
accelerated observers. Such observers are known as Rindler observers. 
We consider Minkowski space with inertial coordinates $(t,x,y,z)$ and
metric $ds^2=-c^2 dt^2+dx^2+dy^2+dz^2$. We perform
a coordinate transformation to new coordinates $(T,Z,X,Y)$
($g>0$ here is the acceleration) 
\begin{equation}
ct=Z \sinh{gT/c},
z= Z \cosh{gT/c},
x=X,
y=Y.
\label{rindler}
\end{equation}
Computing $z^2-c^2t^2=Z^2>0$, we find that this coordinate transformation
only works in the region $|z|^2>c^2t^2$ which consists of the right and left
wedges (Fig.3).  Our interest is only in the right Rindler wedge, where
the Rindler observers are shown in red(Fig.3). 
This transformation is very similar to 
the transformation from Cartesian coordinates in the 
plane to polar coordinates, with $gT/c$ playing the role of the ``angle''
and $Z$ the radial coordinate. Just as circles 
have constant curvature,  the world lines of accelerated observers are
hyperbolae $z^2-c^2 t^2=Z^2=constant$, 
which have constant acceleration. Two of these world lines are shown in red in Fig.3.

There has been much work in quantum field
theory in non-inertial frames (and also in curved spacetimes). A surprising result of this field is that the notion of
a particle is observer dependent\cite{fulling1973}.  
It is known that \cite{davies1975,unruh1976,RevModPhys.80.787}
in the Minkowski vacuum (when the inertial observer sees no particles), 
Rindler observers see a thermal bath of particles with a temperature 
proportional to the acceleration $g$:
\begin{equation}
T= g \frac{\hbar}{2\pi k_B c}.
\label{tempaccnrel}
\end{equation} 
From the analogy with E1, we will readily see that for large enough acceleration, the Unruh photons will deflect the 
interfering electrons and thus destroy the interference pattern.
This physical argument can also be placed on a mathematical footing, by computing the expected value of the Wilson loop in Rindler spacetime.
The calculation is virtually identical to the one in E1 . 
The only difference is that the mode functions are no longer plane waves but those of 
Rindler spacetime. 
In E2, the modes are labelled by $l={\omega,\bf{K}^\perp}$, where $\omega$ is now the frequency as seen by a Rindler observer and $\bf{K}^\perp$ the transverse
wave vector of the mode. These correspond to symmetries of the Rindler spacetime: translation in Rindler time $T$ and the transverse space coordinates
$X,Y$. The modes of Rindler spacetime are plane waves
in the transverse $(X-Y)$ directions and
involve Bessel functions in the $Z$ direction. The formal steps of calculation are exactly the same for both E1 and E2.
Our final conclusion is that acceleration causes destruction of the interference pattern 
in a manner very similar to that shown in Fig.1.

\section{The Gravitational Analogue}
We have focussed our attention on double slit 
experiments with charged particles 
and their decoherence due to fluctuations of the 
electromagnetic field. 
Obviously, this mechanism only works for charged particles.
However, a corresponding mechanism with gravity  
replacing electromagnetism is expected to work for {\it all} particles, 
since gravity is Universal. 
In section II, we had first given a 
physical argument that the interfering  
charged particles scatter  off thermal photons 
and destroy the fringe pattern.
This argument applies also in the gravitational context:
massive (mass $M$) neutral particles scatter 
off thermal gravitons. Within linearised gravity, one ignores
self interaction of the gravitational field, and 
the blackbody gravitons have
a spectrum very similar to photons, as they are both massless
(differing only in spin). 
In a gravitational version of E1, we expect that the thermal bath of gravitons, 
imparts momentum to the interfering
neutral masses and so destroys the interference pattern. 
In E2, we will assume that the perturbatively quantised gravitational field will also lead to 
a thermal bath of Unruh gravitons, just as in the electromagnetic case.

The mathematical analysis is more involved because,
although gravity is a gauge theory like electromagnetism, it is non-Abelian
and therefore nonlinear. 
However, if we confine ourselves to perturbative 
gravity, with $g_{\mu\nu}
=\eta_{\mu\nu}+h_{\mu\nu}$, a linearised
analysis reveals 
that the gravitational Aharonov Bohm effect \cite{fordvilenkin} has a structure
very similar to the electromagnetic case. We have in mind the 
{\it gravielectric}
AB effect \cite{homorgan} rather than the smaller gravimagnetic one.

Imagine an interferometer, rather like LIGO, except that we use
massive particles (as in atom interferometry) rather than photons. 
The action of a neutral particle of mass $M$ traversing a timelike
path $\gamma$ is
\begin{equation}
S=-Mc^2\int (g_{\mu\nu}\frac{ dx^\mu}{d\lambda} \frac{ dx^\nu}{d\lambda})^{1/2}
{d\lambda}  
\label{action}
\end{equation}
leading to a phase $S/\hbar$ in the amplitude for traversing $\gamma$.
Let us first set $g_{\mu\nu}=\eta_{\mu\nu}$ the flat Minkowski metric.
If $\gamma_1$ and $\gamma_2$ are the paths of the two 
arms of the interferometer, there is a fixed phase difference 
between the two arms. If we now perturb the metric away from flat
spacetime: $g_{\mu\nu}=\eta_{\mu\nu}+h_{\mu\nu}$, ($h_{\mu\nu}<<1$) 
as happens for instance when a gravitational wave is incident on the interferometer,
the phase difference depends on the perturbation
$h_{\mu\nu}$, around flat Minkowski space.

We can expand the action
in powers of $h_{\mu\nu}$. The  first term in the expansion 
$S_0$ is $S_0=-Mc^2 \int (\eta_{\mu \nu} u^\mu u^\nu)^{\frac{1}{2}} d\lambda$
(where $u$ is the unit four velocity of the massive particle),
which represents the motion in the flat Minkowski background. 
The first order
correction in $h_{\mu\nu}$ 
is $S_1= -\frac{M c^2}{\hbar} \int h_{\mu\nu} u^\mu u^\nu d\lambda$,
which contributes a gravitational Aharonov-Bohm phase
$\exp{-\frac{M c^2}{\hbar}\int_\gamma h_{\mu\nu} u^\mu u^\nu d \lambda)}$ 
to the interference term. The phase is linear in the perturbation $h_{\mu\nu}$.

We now assume that the perturbation
$h_{\mu\nu}$ is thermal, 
corresponding to a black body radiation of gravitons.
The only difference from the electromagnetic case is that the
thermally fluctuating Wilson loop 
involves $\exp{-iMc^2/2\hbar \int_\gamma h_{\mu\nu} u^\mu u^\nu d \lambda}$ 
(where $u$ is the unit four velocity of the massive particle, normalised
relative to the Minkowski metric)
and therefore, the form factor $\alpha$ is defined accordingly.
The expansion of $h_{\mu\nu}$ in modes and the passage from $E1$ to $E2$
follow the same steps as in electromagnetism. 

As we can expect by analogy, 
the dimensionless fine structure coupling constant 
$e^2/(\hbar c)=1/137$ in (\ref{finalresult}) will be replaced by
$G M^2/(\hbar c)$, leading to 
\begin{equation}
<{\cal W}>=\exp{-\frac{G M^2}{2\hbar c} \sum_{l}  (|\alpha_l|^2 \coth{\frac{\hbar \omega_l}{2 k_B T})}}
\label{finalresult2}
\end{equation} 
which answers exactly to Feynman's expectation 
that the decohering effects will set in when the masses of interfering particles are comparable
to the Planck mass.

\section{Conclusion:}
We have presented a simple solvable model for gravity induced decoherence. Our main results are contained in 
(\ref{finalresult},\ref{finalresult2}) and Fig.2, which show the progressive degradation of coherence.
Our two {\it gedanken} experiments E1 and E2 clarify the relation between acceleration, temperature and decoherence.
The Einstein equivalence principle states that the effects of gravity are indistinguishable from those of acceleration. 
We would therefore conclude from our analysis of E2 that gravity also must have a 
decohering effect on quantum interference. This is the main conclusion of this paper and it is entirely in consonance
with the  proposal of Feynman.  Similar decohering effects are also expected to be seen by 
static observers outside the event horizons of black 
holes due to Hawking radiation.

Before concluding, we address a few questions that 
have been raised by readers of an earlier version of this paper.

{\it Is the Unruh effect real? Is it not too small to observe? 
Do we not need experimental confirmation before we assume that the effect exists?}
Our model for gravity induced decoherence is based on the Unruh effect, 
which is  expected to be present for Rindler observers. 
As noted by Pe\~na and Sudarsky\cite{penasudarsky,vanzellamatsas},
the Unruh effect does not represent new physics beyond that predicted by quantum field
theory in Minkowski space-time. It is merely standard quantum field theory (which has
been rigorously tested in laboratories) viewed from a non-inertial frame. To use an analogy,
if one accepts that Newton's laws of mechanics are valid in inertial frame, one must  
also accept that there will be a Coriolis' effect in a rotating frame. 
Our belief in the Unruh effect need not wait upon further experimental confirmation.
Experiments attempting to see the Unruh effect in acoustic analogues are an ongoing effort.
However, these would only be {\it demonstration} experiments (like the Foucault pendulum in Paris),
illustrating a theoretical prediction  that is already well established.

Like many relativistic effects, the Unruh effect is tiny in normal laboratory
conditions. 
For example, an acceleration equal to the earth's gravity  $g=10\rm{m/s^2}$ corresponds
to a temperature of $10^{-19} \,{}^\circ K$ and a decoherence length of a light
year. For an acceleration equal to the surface gravity of a neutron star,
the temperature is $10^{-9} \,{}^\circ K$ and the decoherence length is $3{\rm Km}$. 
However, it is not the size of the Unruh effect that is the main 
point here. We are concerned with the {\it principle} that gravity
can alter quantum mechanics at large distances. The size of the effects
will increase as one gets to the stronger gravitational fields 
which one expects in the early Universe and the quantum gravity regime.

{\it Would there not be classical radiation from particles at rest (or slow motion) in the accelerating
frame?}\\

This relates to an old discussion about whether a static charge in 
an accelerated frame will radiate. This has been a lively and heated debate and the 
matter has been resolved. Much of the confusion disappears \cite{almeidasaa,eriksengron} when one realises
that the separation of an electromagnetic field into `Coulomb' and `radiation'
fields is frame dependent. A uniformly accelerated charge
in an inertial frame will indeed radiate. One can compute the electromagnetic 
field of the charge using the Lienard-Wiechert potentials. 
However, the {\it same} electromagnetic field if transformed into the frame of a Rindler observer will appear as a pure Coulomb
field due to a charge at rest. In fact, in the Rindler frame there are only
electric fields, no magnetic fields. To summarise, the Rindler observer sees
no radiation, just the Coulomb field of the charge. 
There is no conflict between these two observations. 
The inertial observer has access to spacetime regions not accessible to the Rindler observer.
Within the spacetime regions seen by the Rindler observer, there is a pure Coulomb field.

A similar argument applies to our case where the charges are not static but moving slowly and uniformly.
Since we assume a monoenergetic electron beam, 
the current is steady. In this case there will be electric 
as well as magnetic fields due to the charges, 
but they will all be constant fields, independent of
the Rindler time coordinate in the right Rindler wedge. 
These fields are therefore time independent fields 
and not radiation fields.
In summary, the answer is no, there is no classical radiation in the Rindler frame. The Rindler observer can perform his two slit experiment 
without worrying about radiation from his experiment.

{\it How are divergences in the field correlation function dealt with?}\\
In plotting the curve of Fig. 2, we have plotted the visibility 
relative to its zero temperature value 
$$\frac{{\cal W}(T)}{{\cal W}(0)}.$$ 
This is a vacuum subtraction and the procedure 
is described below. The calculation in the appendix for the loss 
of visibility from a single mode labelled $l$
gives (\ref{bose})
\begin{equation}
{\cal W}_l= \exp{-(n^B_l+1/2) e^2|\alpha_l|^2/2}
\label{renorm}
\end{equation}
Even at zero temperature (when $n^B$ vanishes), there is a residual
term coming from zero point fluctuations.  The ``vacuum subtraction'' 
consists of dropping the $1/2$ in the exponent of (\ref{renorm}). 
In the product over modes (\ref{wilsonproduct}) the term dropped is 
\begin{equation}
\exp{- \sum\limits_{l} \frac{e^2|\alpha_l|^2}{4 \hbar c}}
\label{dropped}
\end{equation}
The sum in the exponent is formally divergent. However, this divergence
comes because, we have pretended that the loop $\gamma$ is 
infinitely thin. This is both physically and mathematically
incorrect, but it simplifies the presentation. 
The loops have to be thickened to many times 
the electron de Broglie wavelength 
to allow for a ``bundle of paths'' as in a real experiment.
In fact, in a real experiment, they would be tubes or wave guides,
which would smear the path $\gamma$ and produce a finite answer.

In E2, we have assumed that the total system is in the 
vacuum of the inertial observer. 
The ``vacuum subtraction'' is done with respect to
the Rindler vacuum.

The simple, solvable 
model for gravity induced decoherence 
proposed here differs considerably from that proposed in
\cite{zych1,zych2}. One way to see this is to note the different
regimes of validity. Our model shows decoherence even when the experiment E2
is done ``horizontally'' (in the $X-Y$ plane, when gravity is along $Z$)
and even when the system in question has no internal structure. In contrast,
the model of Refs\cite{zych1,zych2} {\it need} a composite interfering object
(at least a clock\cite{ss,0264-9381-32-23-239501}, 
which must have at least two internal states) and 
also need a {\it vertical} separation between
parts of the apparatus. More seriously, the effects of \cite{zych1,zych2}
can be undone by reversing the direction of the gravitational field,
as noted by Adler and Bassi\cite{ADLER2016390}. 
The effects we describe in E2 are irreversible. 
In E1, the loss of fringe visibility is due to the fact that
we trace over the photon degrees of freedom, i.e, we do not observe the final
state of the photons. Needless to say, if one works with the total system,
the evolution is still unitary and there is no information 
loss or irreversibility. In E2, however, the final state of the photons 
is inaccessible to the Rindler observer, since they 
are scattered into inaccessible spacetime regions beyond the Rindler horizon. 
The Rindler observer sees irreversible loss of coherence and information. 

Our experiment E1 is closer in spirit to work by Haba \cite{haba1,haba2} and 
Blencowe\cite{blencowe} who also investigate the decohering effect of
thermal gravitons, relics from the early Universe. 
Our results seem to differ in detail from Ref.\cite{blencowe}.
He finds that the decohering effects appear when the energy spread
is of the order of the Planck scale. In our treatment, 
the decohering effect appears even for a monoenergetic system 
($\Delta E=0)$.  Our treatment of E1 is in broad 
agreement with Ref. \cite{haba1,haba2}, but differs in detail 
since Ref. \cite{haba1,haba2} treats a time dependent,
situation while, we are in a simpler stationary one. The 
use of the Unruh effect in E2 to produce a ``thermal 
environment'' from vacuum fluctuations is not considered
in these references.

It will not have escaped the alert reader that while both experiments are proposed as gedanken experiments, 
E1 is well within reach of today's laboratories. 
Apart from the qualitative fact of destruction of interference
fringes, we are also able to quantitatively 
calculate the expected degree of coherence between the two beams. Fig. 1
therefore gives a quantitative prediction which can be tested in the laboratory. The single electron 
experiments \cite{1976AmJPh..44..306M,doi:10.1119/1.16104,1367-2630-15-3-033018}, which have been performed to date use a loop size of order $1{\rm \mu}$.  
At room temperature, decoherence effects are expected to set in when the loop size is about 20 times larger. Of course, the effect
can be enhanced by using charged ions in place of electrons, since the decoherence effect is 
proportional to $e^2$ in the exponent.

The main conclusion of this study is that 
gravity does have a decohering effect on quantum systems,
the effect being larger for sytems which are larger in size and more
strongly coupled. The final formula (\ref{finalresult2})
has the curious feature that it involves the fundamental
constants $\hbar,c,G$ and $k_B$, representing quantum theory, relativity,
gravity and statistical physics.

\section{Acknowledgements:} It is a pleasure to thank T.R. Govindarajan,
H.S. Mani, R. Rajaraman, G. Rajasekharan, Suvrat Raju,
T.R. Ramadas and Supurna Sinha for discussions and Zbigniew Haba for drawing
my attention to \cite{haba1,haba2}. Critical 
comments by two anonymous referees have helped me to improve this paper.
I thank Gee-hyun Yang for spotting a missing factor (eq. 7) 
in the arxiv version, which is fixed the present version.

\section{Appendix}

This appendix presents the main steps involved in the computation of
the thermal average of the Wilson 
loop  ${\cal W}=\exp{ie\int_{\gamma} \bf{A}.d\bf{x}}$. As explained in the text, the
mode decomposition of the $A$ field reduces this to the computation  
of the thermal average of  
${\cal W}_l=\exp{ie(\alpha_l a_l+\alpha^*_l a^\dagger_l)}$,
where $\alpha_l$ is the form factor of the loop $\gamma$ 
and $a_l$ and $a^{\dagger}_l$, the destruction and creation operators
for the $l$th mode. Here we will drop the suffix $l$, it being understood that
$\alpha,\alpha^*,a,a^\dagger, \omega$ and ${\cal W}$, all carry the suffix
$l$. Our notation is standard 
for the quantum harmonic oscillator. In this derivation 
we will set $\hbar,c$ and $k_B$ equal to one and restore them only in the final expression. 
The calculation applies equally to E1 and E2 using the replacement $T=\frac{g}{2\pi}$ to convert the acceleration to the Unruh temperature.

Using the oscillator Hamiltonian $H=(a^\dagger a+1/2)\omega$,
we  find
\begin{equation}
<{\cal W}>= \frac{{\cal N}}{{\cal D}}
\label{nbyd}
\end{equation}
where ${\cal N}=Tr[\rho W]$ and ${\cal D}=Tr[\rho]$, where $\rho=\exp{-\beta (a^{\dagger}a+1/2)\omega}$.
We introduce the short hand notation (used only in this appendix)
$s=e^2|\alpha|^2$ and $t=\exp{-\beta \omega}$. ${\cal D}$ 
evaluates to $\frac{\sqrt{t}}{1-t}$. ${\cal N}$ 
is given by 
\begin{equation}
\label{calneqn}
{\cal N}=\sum\limits_{n=0}^{\infty}\exp{-\beta (n+1/2)\omega} 
\bra{n} {\cal W} \ket{n}.
\end{equation}
The expectation value of ${\cal W}$ is evaluated using 
the standard (Baker-Campbell-Hausdorff) formula
\begin{equation}
{\cal W}=\exp{i e ( \alpha a+ a^\dagger\alpha^* )}=
\exp{-e^2|\alpha|^2/2} \exp{ie\alpha^* a^\dagger}\exp{ie\alpha a},
\label{bch}
\end{equation}
yielding
\begin{equation}
\bra{n} {\cal W} \ket{n}=\exp{-s/2} \bra{n}\exp{i e \alpha^*a^\dagger} \exp{i e\alpha a}\ket{n}.
\label{fff}
\end{equation}
Using the notation $\ket{\psi_n}=\exp{i e\alpha a}\ket{n}$, we find by expanding
the exponential 
\begin{equation}
\label{psin}
\ket{\psi_n}=\sum\limits_{m=0}^n \sqrt{{\binom{n}{m}}}\frac{(i e \alpha)^m}{m!}  \ket{n-m}
\end{equation}
Similarly, letting $\ket{\phi_n}=\exp{-i e\alpha a}\ket{n}$, 
\begin{equation}
\label{phin}
\ket{\phi_n}=\sum\limits_{m=0}^{n}(-1)^m \sqrt{{\binom{n}{m}}}\frac{(i e \alpha)^m}{m!}\ket{n-m}.
\end{equation}
This yields
\begin{equation}
\bra{\phi_n}\ket{\psi_n}= \sum\limits_{m=0}^{n} {\binom{n}{m}} \frac{(-1)^ms^m}{m!}=L_n(s)
\label{phipsi}
\end{equation}
where $L_n(s)$ is the $n$th Laguerre polynomial. 
Using the generating function for the Laguerre polynomials, we get 
\begin{equation}
{\cal N}=\sqrt{t} \exp{-s/2} \sum\limits_{n=0}^{\infty} t^n L_n(s)\\
=\frac{\sqrt{t}}{(1-t)} \exp{-s (\frac{t}{1-t}+1/2)}
\end{equation}
which gives  
\begin{equation}
\frac{\cal N}{\cal D}=\exp{-(n^B+1/2) s}=
\exp{-\frac{e^2|\alpha|^2}{2} \coth{\beta \omega/2}}
\label{bose}
\end{equation}
where $n
^B=t/(1-t)=1/(\exp{\beta \omega}-1)$ is the Bose distribution. 
Restoring the fundamental constants we 
finally arrive at the simple form
\begin{equation}
<{\cal W}_l>=\exp{-\frac{e^2|\alpha_l|^2}{2\hbar c} 
\coth{\frac{\hbar \omega_l}{2k_B T}}}. 
\label{step2}
\end{equation}

\bibliographystyle{utphys}


\begin{thebibliography}{10}

\bibitem{feynman1965flp}
R.~Feynman, R.~Leighton, M.~Sands, and E.~Hafner, {\em {The Feynman Lectures on
  Physics; Vol. III}}, vol.~33.
\newblock AAPT, 1965.

\bibitem{1976AmJPh..44..306M}
P.~G. {Merli}, G.~F. {Missiroli}, and G.~{Pozzi}, ``{On the statistical aspect
  of electron interference phenomena},''
  \href{http://dx.doi.org/10.1119/1.10184}{{\em American Journal of Physics}
  {\bfseries 44} (Mar., 1976) 306--307}.

\bibitem{doi:10.1119/1.16104}
A.~Tonomura, J.~Endo, T.~Matsuda, T.~Kawasaki, and H.~Ezawa, ``Demonstration of
  single electron buildup of an interference pattern,''
  \href{http://dx.doi.org/10.1119/1.16104}{{\em American Journal of Physics}
  {\bfseries 57} no.~2, (1989) 117--120},
  \href{http://arxiv.org/abs/http://dx.doi.org/10.1119/1.16104}{{\ttfamily
  http://dx.doi.org/10.1119/1.16104}}. \url{http://dx.doi.org/10.1119/1.16104}.

\bibitem{1367-2630-15-3-033018}
R.~Bach, D.~Pope, S.-H. Liou, and H.~Batelaan, ``Controlled double-slit
  electron diffraction,'' {\em New Journal of Physics} {\bfseries 15} no.~3,
  (2013) 033018. \url{http://stacks.iop.org/1367-2630/15/i=3/a=033018}.

\bibitem{diosi1987universal}
L.~Diosi, ``A universal master equation for the gravitational violation of
  quantum mechanics,'' {\em Physics letters A} {\bfseries 120} no.~8, (1987)
  377--381.

\bibitem{diosi1989models}
L.~Diosi, ``Models for universal reduction of macroscopic quantum
  fluctuations,'' {\em Physical Review A} {\bfseries 40} no.~3, (1989) 1165.

\bibitem{penrose}
R.~Penrose, ``On gravity's role in quantum state reduction,'' {\em General
  Relativity and Gravitation} {\bfseries 28} (1996) 581.

\bibitem{feynmanchapelhill}
D.~Rickles and C.~DeWitt, {\em The Role of Gravitation in Physics}.
\newblock Max Planck Institute, 2011.
\newblock \url{http://www.edition-open-sources.org/sources/5/index.html}.

\bibitem{feynman2002feynman}
R.~Feynman, F.~Morinigo, W.~Wagner, and B.~Hatfield, {\em Feynman Lectures on
  Gravitation}.
\newblock Frontiers in Physics Series. Avalon Publishing, 2002.
\newblock \url{https://books.google.co.in/books?id=jL9reHGIcMgC}.

\bibitem{penrose2014}
R.~Penrose, ``On the gravitization of quantum mechanics 1: Quantum state
  reduction,'' \href{http://dx.doi.org/10.1007/s10701-013-9770-0}{{\em
  Foundations of Physics} {\bfseries 44} no.~5, (2014) 557--575}.
  \url{http://dx.doi.org/10.1007/s10701-013-9770-0}.

\bibitem{zych1}
M.~Zych, F.~Costa, I.~Pikovski, and C.~Brukner, ``Quantum interferometric
  visibility as a witness of general relativistic proper time,'' {\em Nat
  Commun} {\bfseries 2} (Oct, 2011) 505.
  \url{http://dx.doi.org/10.1038/ncomms1498}.

\bibitem{zych2}
I.~Pikovski, M.~Zych, F.~Costa, , and C.~Brukner, ``Universal decoherence due
  to gravitational time dilation,'' \href{http://dx.doi.org/DOI:
  10.1038/ncomms1498}{{\em Nat Phys} (June, 2015) 668}.
  \url{http://arxiv.org/abs/1311.1095}.

\bibitem{ADLER2016390}
S.~L. Adler and A.~Bassi, ``Gravitational decoherence for mesoscopic systems,''
  \href{http://dx.doi.org/http://dx.doi.org/10.1016/j.physleta.2015.10.064}{{\em
  Physics Letters A} {\bfseries 380} no.~3, (2016) 390 -- 393}.
  \url{http://www.sciencedirect.com/science/article/pii/S0375960115009561}.

\bibitem{diosi2015}
L.~Diosi, ``{Centre of mass decoherence due to time dilation: paradoxical
  frame-dependence},''
\href{http://arxiv.org/abs/1507.05828}{{\ttfamily arXiv:1507.05828
  [quant-ph]}}.

\bibitem{minarPhysRevA.94.062111}
J.~c.~v. Min\'a\ifmmode~\check{r}\else \v{r}\fi{}, P.~Sekatski, and
  N.~Sangouard, ``Bounding quantum-gravity-inspired decoherence using atom
  interferometry,'' \href{http://dx.doi.org/10.1103/PhysRevA.94.062111}{{\em
  Phys. Rev. A} {\bfseries 94} (Dec, 2016) 062111}.
  \url{https://link.aps.org/doi/10.1103/PhysRevA.94.062111}.

\bibitem{fulling1973}
S.~A. {Fulling}, ``{Nonuniqueness of Canonical Field Quantization in Riemannian
  Space-Time},'' \href{http://dx.doi.org/10.1103/PhysRevD.7.2850}{{\em Phys.
  Rev. D} {\bfseries 7} (May, 1973) 2850--2862}.

\bibitem{davies1975}
P.~C.~W. {Davies}, ``{Scalar production in Schwarzschild and Rindler
  metrics},'' \href{http://dx.doi.org/10.1088/0305-4470/8/4/022}{{\em Journal
  of Physics A Mathematical General} {\bfseries 8} (Apr., 1975) 609--616}.

\bibitem{unruh1976}
W.~G. {Unruh}, ``{Notes on black-hole evaporation},''
  \href{http://dx.doi.org/10.1103/PhysRevD.14.870}{{\em Phys. Rev. D}
  {\bfseries 14} (Aug., 1976) 870--892}.

\bibitem{RevModPhys.80.787}
L.~C.~B. Crispino, A.~Higuchi, and G.~E.~A. Matsas, ``The unruh effect and its
  applications,'' \href{http://dx.doi.org/10.1103/RevModPhys.80.787}{{\em Rev.
  Mod. Phys.} {\bfseries 80} (Jul, 2008) 787--838}.
  \url{https://link.aps.org/doi/10.1103/RevModPhys.80.787}.

\bibitem{fordvilenkin}
L.~H. Ford and A.~Vilenkin, ``A gravitational analogue of the aharonov-bohm
  effect,'' {\em Journal of Physics A: Mathematical and General} {\bfseries 14}
  no.~9, (1981) 2353. \url{http://stacks.iop.org/0305-4470/14/i=9/a=030}.

\bibitem{homorgan}
V.~B. Ho and M.~J. Morgan, ``An experiment to test the gravitational
  aharonov-bohm effect,'' {\em Aust. J. Phys} {\bfseries 47} (1994) 245.

\bibitem{penasudarsky}
I.~Pena and D.~Sudarsky, ``On the possibility of measuring the unruh effect,''
  {\em Foundations of Physics} {\bfseries 44} (2014) 689--708.

\bibitem{vanzellamatsas}
D.~A.~T. Vanzella and G.~E.~A. Matsas, ``Decay of accelerated protons and the
  existence of the fulling-davies-unruh effect,''
  \href{http://dx.doi.org/10.1103/PhysRevLett.87.151301}{{\em Phys. Rev. Lett.}
  {\bfseries 87} (Sep, 2001) 151301}.
  \url{https://link.aps.org/doi/10.1103/PhysRevLett.87.151301}.

\bibitem{almeidasaa}
C.~de~Almeida and A.~Saa, ``The radiation of a uniformly accelerated charge is
  beyond the horizon: A simple derivation,''
  \href{http://dx.doi.org/10.1119/1.2162548}{{\em American Journal of Physics}
  {\bfseries 74} no.~2, (2006) 154--158},
  \href{http://arxiv.org/abs/https://doi.org/10.1119/1.2162548}{{\ttfamily
  https://doi.org/10.1119/1.2162548}}. \url{https://doi.org/10.1119/1.2162548}.

\bibitem{eriksengron}
E.~Eriksen and A.~Gron, ``Electrodynamics of hyperbolically accelerated
  charges,''
  \href{http://dx.doi.org/https://doi.org/10.1006/aphy.2000.6096}{{\em Annals
  of Physics} {\bfseries 286} no.~2, (2000) 373 -- 399}.
  \url{http://www.sciencedirect.com/science/article/pii/S0003491600960962}.

\bibitem{ss}
S.~Sinha and J.~Samuel, ``Atom interferometry and the gravitational redshift,''
  {\em Classical and Quantum Gravity} {\bfseries 28} no.~14, (2011) 145018.

\bibitem{0264-9381-32-23-239501}
S.~Sinha and J.~Samuel, ``Erratum and addendum: Atom interferometry and the
  gravitational redshift (2011 class. quantum grav. 28
  [http://dx.doi.org/10.1088/0264-9381/28/14/145018] 145018 ),'' {\em Classical
  and Quantum Gravity} {\bfseries 32} no.~23, (2015) 239501.
  \url{http://stacks.iop.org/0264-9381/32/i=23/a=239501}.

\bibitem{haba1}
Z.~Haba, ``Decoherence by relic gravitons,'' {\em Phys. Lett.} {\bfseries A 15}
  (2000) 1519.

\bibitem{haba2}
Z.~Haba, ``Gravitational decoherence and epr correlations,'' {\em Int. Journal
  of Theoretical Physics} {\bfseries 40} (2001) 985.

\bibitem{blencowe}
M.~P. Blencowe, ``Effective field theory approach to gravitationally induced
  decoherence,'' \href{http://dx.doi.org/10.1103/PhysRevLett.111.021302}{{\em
  Phys. Rev. Lett.} {\bfseries 111} (Jul, 2013) 021302}.
  \url{https://link.aps.org/doi/10.1103/PhysRevLett.111.021302}.

\end{thebibliography}
\providecommand{\href}[2]{#2}\begingroup\raggedright\endgroup

\providecommand{\href}[2]{#2}\begingroup\raggedright

\endgroup

\end{document}